\documentstyle[psfig,twocolumn,aps,pre]{revtex}

\psfigurepath{figs}

\def \lanl#1{{\tt (cond-mat/#1)}}

\def \prl#1#2#3{{ Phys.~Rev.~Lett. {\bf #1} (#2), #3}}
\def \pre#1#2#3{{ Phys.~Rev.~E {\bf #1} (#2), #3}}
\def \prb#1#2#3{{ Phys.~Rev.~B {\bf #1} (#2), #3}}

\def \sciam#1#2#3{{ Scientific American, #1 issue #2, p. #3}}
\def \eurpL#1#2#3{{ Europhys.~Lett. {\bf #1} (#2), #3}}
\def \jpa#1#2#3{{ J.~Phys.~A: Math.~Gen. {\bf #1} (#2), #3}}
\def \jpc#1#2#3{{ J.~Phys.~C: Solid State {\bf #1} (#2), #3}}

\def \repp#1#2#3{{ Rep.~Prog.~Phys. {\bf #1} (#2), #3}}

\def \jpq1#1#2#3{{ J.~Physique {\it I} (France) {\bf #1} (#2), #3}}
\def \jpqL#1#2#3{{ J.~Physique Lett.~(France) {\bf #1} (#2), #3}}
\def \jsp#1#2#3{{  J.~Stat.~Phys. {\bf #1} (#2), #3}}

\def \jcompp#1#2#3{{  J.~Comp.~Phys. {\bf #1} (#2), #3}}

\def \IJMPC{{  Int.~J.~Mod.~Phys.~C}}

\def \siamjc#1#2#3{{ SIAM J.~Comput. {\bf #1}(#2), #3}}

\begin{document}
\title{Comparison of rigidity and connectivity
percolation in two dimensions}
\author{C.~Moukarzel\footnote{ \noindent {\bf email:} cristian@if.uff.br}}
\address{Instituto de F\'\i sica, Universidade Federal Fluminense, \\
CEP 24210-340, Niteroi RJ, Brazil.}
\author{P.~M.~Duxbury}
\address{ Dept. of Physics/Ast. and Center for Fundamental Materials
  Research,\\ 
  Michigan State University, East Lansing, MI 48824, USA.  } 
\maketitle
\begin{abstract}
  
  Using a recently developed algorithm for generic rigidity of two-dimensional
  graphs, we analyze rigidity and connectivity percolation transitions in two
  dimensions on lattices of linear size up to $L=4096$. We compare three
  different universality classes: The generic rigidity class; the connectivity
  class and; the generic ``braced square net''(GBSN).  We analyze the spanning
  cluster density $P_\infty$, the backbone density $P_B$ and the density of
  dangling ends $P_D$.  In the generic rigidity and connectivity cases, the
  load-carrying component of the spanning cluster, the backbone, is
  \emph{fractal} at $p_c$, so that the backbone density behaves as $B \sim
  (p-p_c)^{\beta'}$ for $p > p_c$. We estimate $\beta'_{gr} = 0.25 \pm 0.02 $
  for generic rigidity and $\beta'_c=0.467\pm 0.007$ for the connectivity
  case.  We find the correlation length exponents, $\nu_{gr} =1.16\pm 0.03$
  for generic rigidity compared to the exact value for connectivity
  $\nu_c=4/3$.  In contrast the GBSN undergoes a first-order rigidity
  transition, with the backbone density being extensive at $p_c$, and
  undergoing a jump discontinuity on reducing $p$ across the transition.  We
  define a model which tunes continuously between the GBSN and GR classes and
  show that the GR class is typical.

\end{abstract}
\section{ Introduction }
\label{sec:intro}

Scalar Percolation is a simple model for disordered systems, and has received
much attention in the last two decades~\cite{Stf_bk,Ess_rv,BuHa}.  This model
describes the transmission of a scalar conserved quantity (for example
electric charge, or fluid mass) across a randomly diluted system.  However in
the calculation of mechanical properties force ( i.e. a vector) must be
transmitted across the system~\cite{Sah_rv}. It was originally
suggested\cite{deGe} that the critical behavior of the elastic moduli of a
percolating system should be equivalent to that of its conductivity, but this
is only valid for the scalar limit of the Born model of elasticity\cite{Born},
a model which is not rotationally invariant and in many cases inappropriate.
Elastic percolation is not in general equivalent to scalar percolation. This
was first made clear by the work of Feng and Sen\cite{Fng1}, which showed that
{\it central-force elasticity} percolation was in a different universality
class than scalar percolation, and provided the starting point for a renewed
interest in this problem.

It soon became clear that the elasticity problem can be divided in two
categories\cite{Th1}, according to the kind of forces which hold the lattice
together. If angular forces are important\cite{Kan,Fng2,Fng3,Shm1,ZaSt,Rx1}, a
singly-connected path across the lattice is enough to ensure rigidity, so any
configuration of bonds which is connected is also rigid. In this case, the
geometry of the elastic backbone is exactly the same as that of the scalar
percolation problem \cite{Fng2,Shm1,ZaSt,Rx1}.  This is the case for
bond-bending\cite{Kan,Fng2} and beam\cite{Fng3,Rx1} models.  Thus, the
elasticity percolation problem with angular forces is now well understood and
that understanding has borrowed much from the \emph{geometrically equivalent}
scalar percolation problem. It is the purpose of this paper to analyze the
more challenging central-force rigidity percolation transition, stressing the
similarities and differences between this problem and scalar percolation.

\begin{figure}[htb] \vbox{ 
\centerline{ \psfig{figure=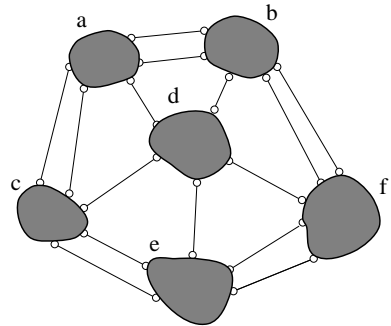,width=5cm,angle=270} }
\centerline{}
\caption{ 
The six bodies shown in this figure are rigidly connected, i.e. they
belong to the same rigid cluster. But the removal of any bond (thin
black lines)  leads to the collapse of the structure, which is then
reduced to a collection of six \emph{independent} rigid clusters (no two
are rigidly connected). This means that the existence of a rigid
connection between for example clusters $e$ and $f$ cannot be decided on
local information only, since it depends on the presence of 'far away'
bonds, i.e. bonds not connected to clusters $e$ or $f$.
}
\label{fig:collapse}
} \end{figure}

If rigidity~\cite{Fng1,DaTr,RH0,RH2,KnSah,ArSah,MDlet1,MDlet2,JaTh,Guy,Obkv}
is provided by central forces alone (e.g.  rotatable springs),
single-connectedness is not enough to ensure rigidity. In this case a lattice
that is conducting usually would not support an applied stress (i.e.  it would
not be {\it rigid}). This was first shown by Feng and Sen~\cite{Fng1} who
found that the rigidity threshold is significantly larger than the
conductivity threshold. An exception worth mentioning is the case of elastic
lattices under tension, or equivalently, systems in which all springs have
zero repose length~\cite{Tang}. For such systems, conductivity and Young
modulus are \emph{equivalent}, i.e. rigidity appears at the scalar percolation
point.  It has been recently emphasized~\cite{Ent} that entropic effects can
give rise to similar effects in central-force systems with nonzero repose
length and finite temperature, although the connection with ref.~\cite{Tang}
was not established.

The main difficulties associated with central-force elasticity are as follows:
Whereas in the scalar connectivity case it is a trivial problem to determine
when two sites belong to the same connected cluster, in the case of
central-force rigidity it is not in general easy to decide whether two objects
are rigidly connected. For example it takes some thinking to see that the six
bodies in Fig.~\ref{fig:collapse} form a rigid unit. Thus it is not easy
to see how a computer algorithm can be devised to identify rigid clusters.

In the scalar connectivity problem, the removal of a singly-connected bond
leads to the separation of a connected cluster into two clusters.  In the
rigidity case, the removal of an analogous ``cutting bond'' may produce the
collapse of a rigid cluster to a collection of an arbitrarily large number of
smaller ones (we call this the {\it house-of-cards} mechanism).
Fig.~\ref{fig:collapse} shows a simple example of this situation. Due to this
fact, the transmission of rigidity can be ``non-local'' \cite{DaTr,Guy}, since
a bond added between two clusters on one side of the sample may induce
rigidity between two clusters on the other side of a sample. 

\begin{figure}[htb] \vbox{ 
\centerline{ \psfig{figure=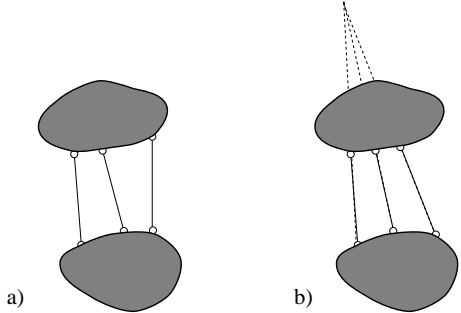,width=6cm,angle=270} }
\vskip 1cm
\caption{ 
  Three bars are in general enough to form a rigid connection between two
  rigid bodies (case {\bf a}), but for particular, \emph{degenerate} cases
  (case {\bf b}), rigidity fails even when the system has the right number of
  bars. Case {\bf b} fails to be rigid because the three bars happen to have a
  common point.  A structure formed by two bodies connected by three bars is
  \emph{generically rigid} in two dimensions if it is rigid ``for most
  geometrical arrangements'', i.e. leaving aside degenerate configurations
  such as {\bf b)}, which occur with probability zero in the configuration
  space.  }
\label{fig:degeneracy}
} \end{figure}

A second source of difficulties in the problem of central-force rigidity is
the existence of particular geometrical arrangements for which a system may
fail to be rigid~\cite{infrig} even if it is rigid for most other
cases\cite{Rig_def,Hendr,CFMalg,Guy}. Take for example two rigid bodies
connected by three bars in two dimensions, as shown in
Fig.~\ref{fig:degeneracy}. This structure is in general rigid, but if the
three bars happen to have a common point, then the structure is not rigid,
since this common point is the center of relative rotations\cite{Guy} between
the two bodies.

Particular geometrical arrangements (such as Fig.~\ref{fig:degeneracy}b),
which are non-rigid even when the structure is rigid for most other
configurations, are called \emph{degenerate configurations}. Degenerate
configurations appear with probability zero if the lattice locations are
``randomly chosen''.  A lattice is thus said to be \emph{generically rigid},
if it is rigid for most geometrical arrangements of its sites. Generic
rigidity only depends on the topology of the underlying graph, i.e. ignores
the possibility of degeneracies. Since degenerate configurations are always
possible on perfectly regular lattices, we will assume our lattice sites to
have small random displacements, in which case generic rigidity applies.

Up to recently, no simple algorithms existed for the determination of the
rigid-cluster structure of arbitrary lattices.  Due to this, direct solving of
the elastic equations was one of the few methods~\cite{Sah_rv} available to
obtain information on the structure of rigidly connected clusters. But this is
very time-consuming and did not allow the study of large lattices. Previous
simulations were for example not sufficiently precise to confirm or reject the
proposal\cite{RH0,RH1,RH2} that bond-bending and central-force elastic
percolation might after all still be in the same universality class. This
suggestion was not inconsistent with some numerical results obtained on
small-sized systems\cite{RH0,RH1,RH2,KnSah,ArSah}.

Recently there has been a breakthrough in the system sizes accessible to
numerical analysis\cite{MDlet1,JaTh,MDlet2}, following the development of
efficient graph-theoretic methods for the problem or generic rigidity in two
dimensions\cite{Hendr,CFMalg,PG}. Using such methods, we study the
central-force rigidity percolation problem on randomly diluted triangular
lattices of linear size up to $L=3200$, and connectivity percolation and
body-joint rigidity percolation on site-diluted square lattices of size up to
$L=4096$.  Our numerical algorithm~\cite{CFMalg} is complementary to the
``pebble game''~\cite{JaTh,PG}, which is an implementation of Hendrickson's
matching algorithm in the original ``joint-bar'' representation of the
network~\cite{Hendr}(see below).  This paper is an elaboration and extension
of our two recent letters on this subject\cite{MDlet1,MDlet2}.  We extend and
elaborate upon the numerical data presented there in several ways: by
comparing rigidity and connectivity percolation, by studying significantly
larger lattices for rigidity percolation, by giving data on site and bond
diluted lattices with a variety of boundary conditions; by presenting results
on a new body-joint model which is in the universality class of bar-and-joint
rigidity percolation and; by presenting detailed results for a model which
continuously tunes between the braced square lattice (which has a first order
rigidity transition) and the isotropic triangular lattice (which has a second
order rigidity transition).

The numerical method is briefly described in Section~\ref{sec:method}. In
Section~\ref{sec:results}, our results are presented and their implications
discussed. A comparison is made with other available numerical and analytical
results for the central-force rigidity percolation problem. We also discuss
the issue of first-order rigidity, which has been the subject of a comment and
reply in physical review letters\cite{JTcom}. Section \ref{sec:conclusions}
contains our conclusions.

\section{The numerical method}
\label{sec:method}

We take an initially depleted triangular lattice and add bonds (in the
bond-diluted case) or sites (in the site-diluted case) to it one at a time,
and use a graph-theoretic matching algorithm~\cite{CFMalg} in order to
identify the rigid clusters that are formed in the system.  For the case of
bond dilution, $p$ is the density of present bonds, while in site dilution it
indicates the density of present sites. In the site-diluted problem, a bond is
present if the two sites it connects are.

We use the body-bar version~\cite{CFMalg} of a recently proposed rigidity
algorithm~\cite{Hendr}. This algorithm, being combinatorial in nature, allows
us to identify sets of sites which are rigidly connected, without providing
any information on the actual values of the stresses when an external load is
applied. The body-bar algorithm sees the lattice as a collection of rigid
clusters (or ``bodies'') connected by bars, instead of points connected by
bars as proposed in the original algorithm\cite{Hendr} and as implemented in
the ``pebble game''\cite{PG}. The body-bar representation allows a more
efficient use of CPU and memory, as each rigid cluster is represented as one
object. The matching identifies rigid clusters and \emph{condenses} them to
one node as new bonds are added to the network.

In two-dimensional rigidity, a rigid cluster has $3$ degrees of freedom, while
a point-like joint has $2$. Therefore the \emph{minimum} number of bonds
needed to rigidize $n$ joints in 2d is $(2n-3)$. Matching
algorithms~\cite{Hendr,CFMalg,PG} are based on this sort of
constraint-counting.

\begin{figure}[htb]
\vbox{
\centerline{\psfig{figure=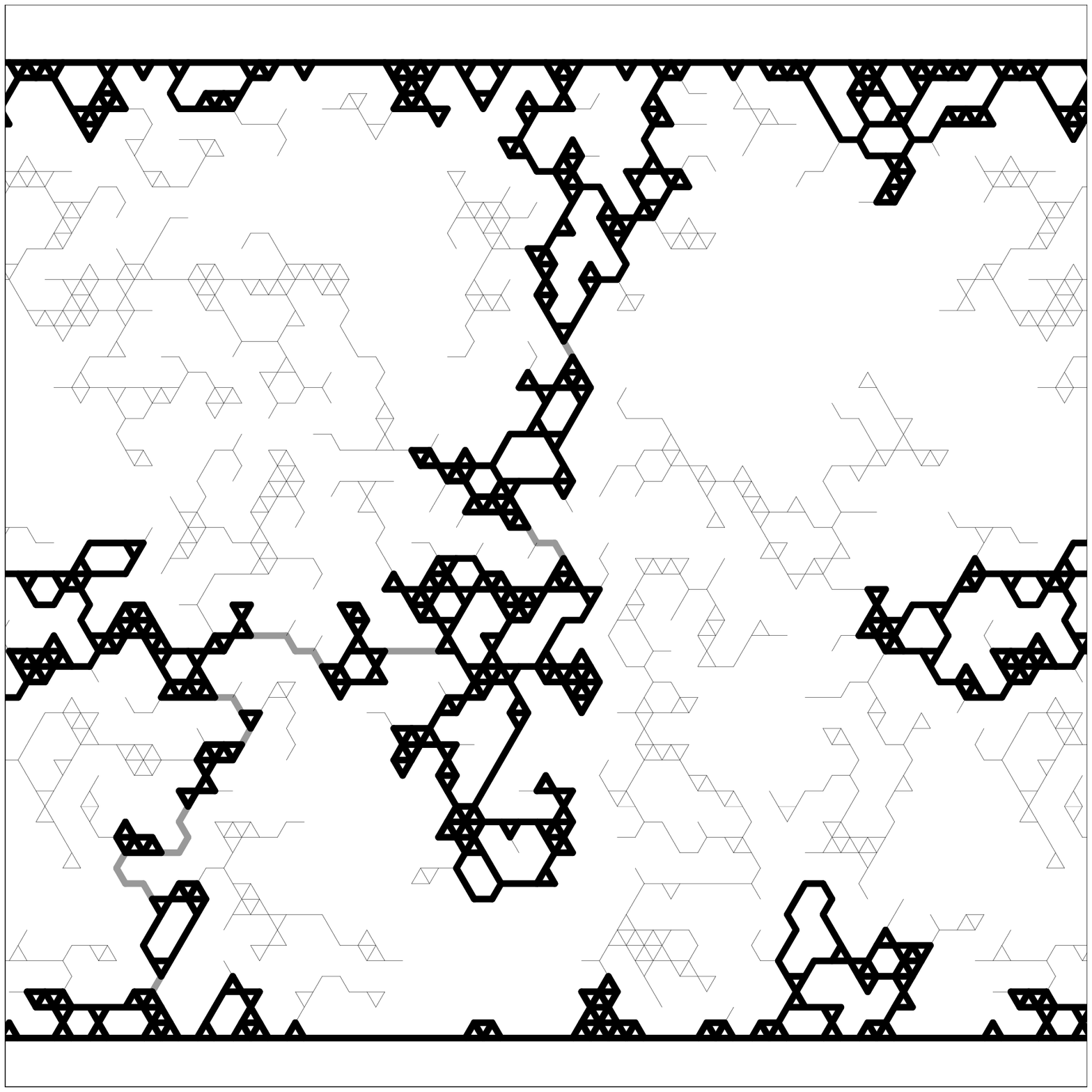,width=6cm}}
\centerline{\psfig{figure=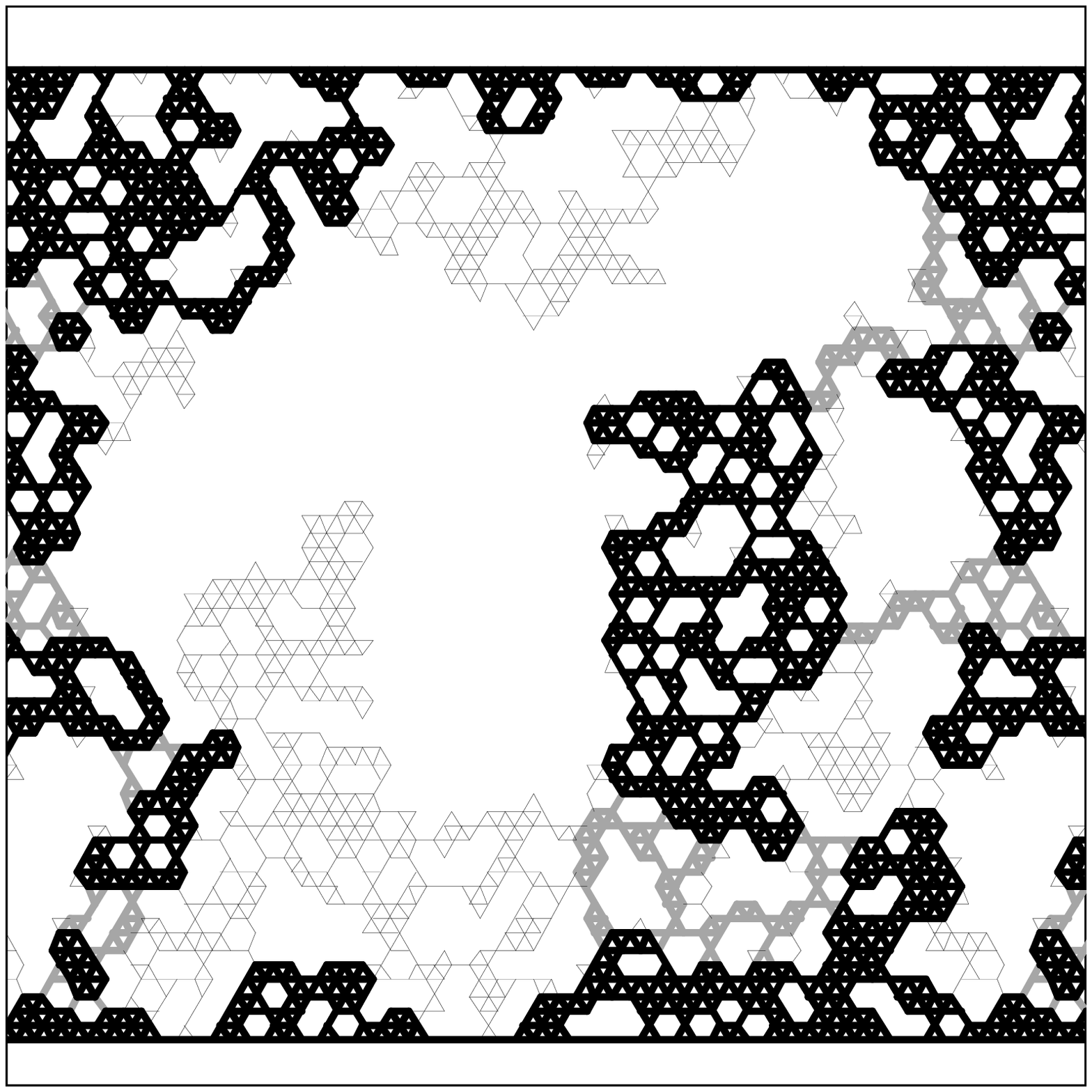,width=6cm}}  
\centerline{\psfig{figure=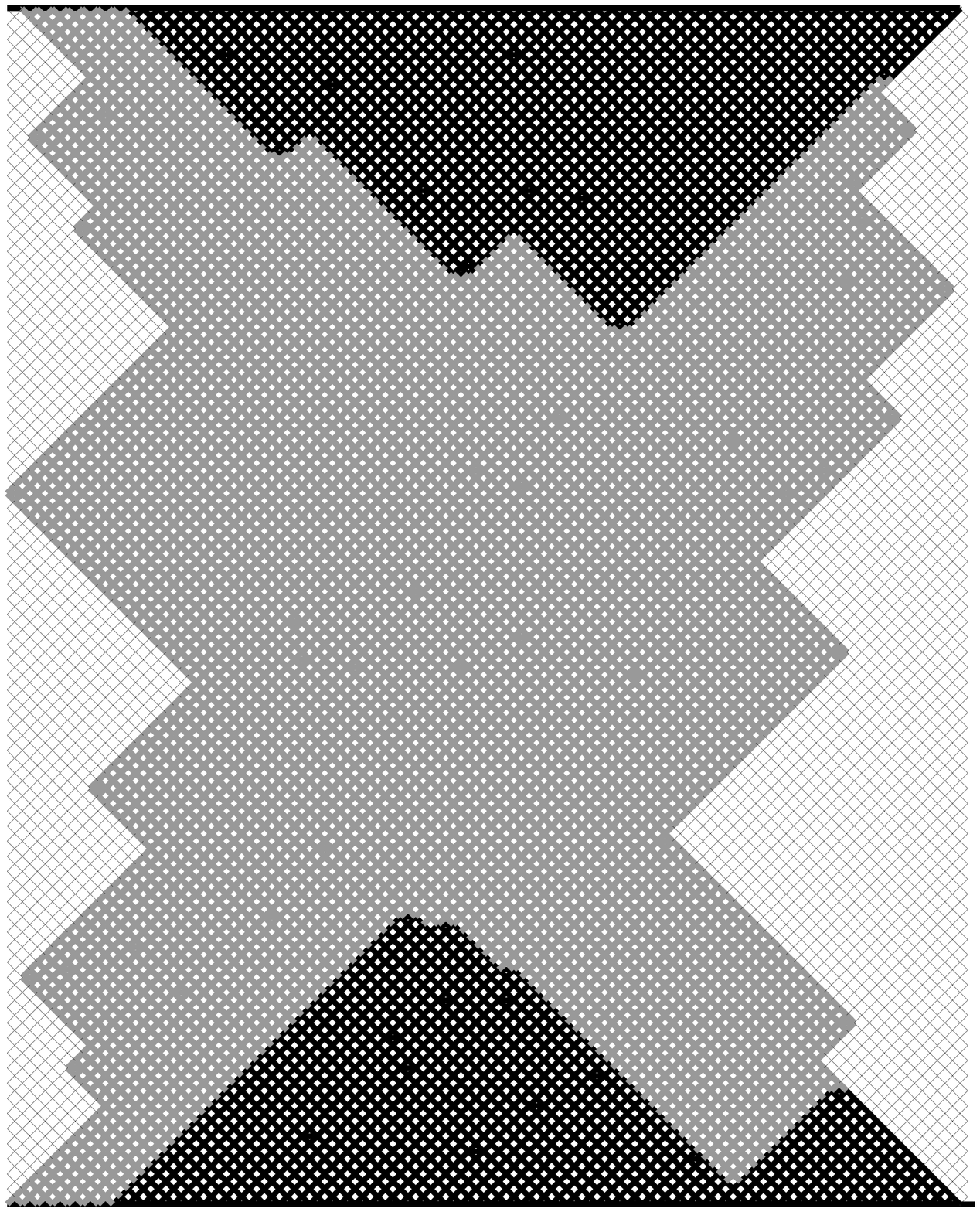,height=5.8cm,width=6cm}}}
\vskip 0.2cm
\caption{
  Infinite percolation clusters which lie in different universality classes:
  {\bf a)} Connectivity percolation ($g=G=1$) on a triangular lattice; {\bf
    b)} Rigidity percolation ($g=2, G=3$) on a triangular lattice; {\bf c)}
  Rigidity percolation ($g=2, G=3$) on a braced square lattice.  For a) and
  b), boundary conditions are periodic in the horizontal direction while for
  c) they are free. The system size $L=64$ and rigid bus-bars are set on the
  upper and lower ends of the sample. The \emph{ backbone}, is composed of
  'blobs' of internally stressed bonds (thick black lines), rigidly
  interconnected by cutting bonds (gray lines). Cutting bonds are also called
  \emph{red bonds}. Removing one of them produces the collapse of the system.
  \emph{Dangling ends} (thin lines) are rigidly connected to the backbone, but
  do not add to the ability of these networks to carry a DC external load (or
  current). }
\label{fig:spcluster}
\end{figure}

The body-bar algorithm~\cite{CFMalg}, can be extended to handle ``rigidity
problems'' with arbitrary values of $g$ (number of degrees of freedom of a
joint) and $G$ (degrees of freedom of a rigid cluster).  Connectivity for
example, is just a special (simple) case of rigidity with $g=1$ and $G=1$: the
minimum number of bonds needed to connect $n$ points is $n-1$ in \emph{any}
dimension. Connectivity percolation can thus be studied using this algorithm.
More details on the application of matching algorithms for the specific case
of connectivity percolation can be found in Ref.~\cite{CFMbackbone}.

There are several ways to define the onset of global rigidity in a
network~\cite{MDlet1}.  We have used two distinct methods.  First we determine
whether an externally applied stress can be supported by the network, which we
call applied stress (AS) percolation.  Secondly we studied the percolation of
internally-stressed (IS) regions.  

At the AS \emph{percolation point}, an applied stress is first able to be
transmitted between the lower and upper sides of the sample. As we add bonds
one at a time, we are able to exactly detect this percolation point by
performing a simple test~\cite{CFMalg} which consists in connecting an
additional fictitious spring between the upper and lower sides of the system.
This auxiliary spring mimics the effect of an external load, and therefore the
first time that a macroscopic rigid connection exists, a globally stressed
region (the backbone) appears.

The IS critical point is defined as the bond- or site density at which
internal stresses percolate through the system.  This means that the upper and
lower sides of the system belong to the same self-stressed
cluster\cite{MDlet1}, and this is trivially detected within the matching
algorithm\cite{CFMalg}.  The AS and IS definitions of percolation are in
principle different, but we found~\cite{MDlet1} that the average percolation
threshold and the critical indices coincide for large lattices.  Similar
definitions apply to the connectivity case, with the AS case being the usual
definition, i.e. the onset of electric conductivity, and IS being percolation
of ``Eddy currents''.

We define the \emph{spanning cluster} (Fig.~\ref{fig:spcluster}) as
the set of bonds that are rigidly connected to both sides of the sample.
However only a subset of these bonds carry the applied load.  This subset is
called the \emph{backbone}.  The backbone will in general include some
\emph{cutting bonds}, so named because the removal of any one of them leads to
the loss of global load carrying capability. Cutting bonds attain their
maximum number exactly at $p_c$~\cite{Coniglio}. The backbone bonds which are
not cutting bonds are parts of internally overconstrained \emph{blobs}. In the
rigidity case, the smallest overconstrained cluster on a triangular lattice is
the complete hexagonal wheel (twelve bonds), while in the connectivity case it
is a triangle (i.e.  the smallest possible \emph{loop}). The spanning cluster
also contains bonds which are rigidly connected to both ends of the sample but
which do not carry any of the applied load. These are called \emph{dangling
  ends}. This classification is standard in connectivity (scalar)
percolation~\cite{Stanley}. 

In this work we analyze several other boundary conditions, particularly in the
generic rigidity case on triangular lattice.  In that case, for site dilution
we analyze: AS with rigid bus-bars at the ends of the sample, AS without bus
bars (any-pair rigidity), and IS with bus-bars.  For bond dilution, only the
AS with bus-bars case was studied.  We determine the exact percolation point
(AS or IS) for each sample, so we can identify and measure the different
components of the spanning cluster {\it exactly} at $p_c$ for each sample.
This should be contrasted with usual numerical approaches, in which averages
are done at fixed values of $p$, and $<p_c>$ is obtained from finite-size
scaling (e.g. data collapse). In that case, it is known that slight
differences in the estimation of $<p_c>$ can lead to important deviations in
critical indices~\cite{RH0}. This source of error is absent in our
measurements.  Sample averages are done over approximately $10^8/L^2$ samples.

\section{ Results }
\label{sec:results}

We first analyze the size dependence of the three key probabilities $P_B$, the
backbone density, $P_D$, the dangling end density and $P_{\infty}$, the
infinite cluster density, {\it exactly at their percolation thresholds} as
described in the previous section. In Fig. \ref{fig:components-all}a-c, this
data is presented for the three different universality classes in Figs.
\ref{fig:spcluster}a-c.

\vbox{
\begin{figure}[htb]
\centerline{\psfig{figure=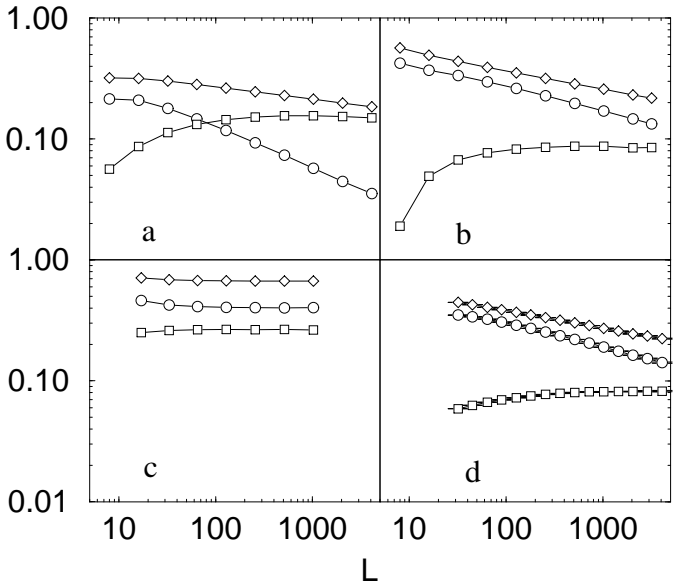,width=9cm,angle=270}}
\vskip 1cm
\caption{ 
Density of backbone bonds (circles), dangling bonds (squares) and infinite 
cluster bonds (diamonds) at the AS critical point for: {\bf a)} Connectivity
percolation 
($g=G=1$) on a site-diluted square lattice; {\bf b)} Rigidity 
percolation ($g=2, G=3$) on a site-diluted triangular lattice; 
 {\bf c)} Rigidity on a randomly braced square lattice;  {\bf d)} Body
rigidity ($G=g=3$) on a site-diluted square lattice . 
}
\label{fig:components-all}
\end{figure}
}

Case c corresponds to the generic braced square net (GBSN), which is a square
lattice to which diagonals are added at random with probability $p$. The
non-generic version of this problem has been studied by many
authors~\cite{BC,Dewney,Obkv}, and it is well known that the number of
diagonals needed to rigidize it is not extensive: $p_c \sim 0$ when $L\to
\infty$. This is confirmed by our numerical simulations, which correspond to
the bus-bar boundary condition.

In Fig.~\ref{fig:components-all}d we also present data for the rigidity case
$g=3,G=3$ on a square lattice, to further test whether the rigidity class is
universal in two dimensions.  In this model each site of a square lattice is a
body and so has $g=3$ degrees of freedom.  Each of these bodies is connected
to each adjacent body by two bonds or bars, i.e. two contiguous bodies are
pinned at a common point.  Maxwell counting~\cite{Maxap} then implies \hbox{$f
  = 3 - 4 p$}, so that the Maxwell estimate of the bond percolation threshold
is $3/4$. Our numerical estimate is $p_c=0.74877 \pm 0.00005$, thus confirming
the accuracy of the Maxwell approximation.

One clear feature of Fig.~\ref{fig:components-all} is that the BSN
(Fig.~\ref{fig:components-all}c) has a qualitatively different behavior than
the other cases.  For the BSN, $P_B$, $P_\infty$ and $P_D$ all have a finite
density at large $L$, indicating that the rigidity transition is first order
in this case.  In contrast, in both the
connectivity(Fig.~\ref{fig:components-all}a) and rigidity cases (Figs.
\ref{fig:components-all}b,d), $P_B$ and $P_{\infty}$ are decreasing in a power
law fashion over the available size ranges.  However the behavior of $P_D$ is
more complex.  First we discuss the behavior of $P_B$.
 
At a second order phase transition, finite-size scaling theory predicts $P_B
(p_c) \sim L^{-\beta'/\nu}$. Taking into account correction-to-scaling terms,
which we assume to be power-law, we may generally write

\begin{equation}
P_B = C_1  L^{-e} (1 + C_2 L^{-\omega}).
\label{eq:fit}
\end{equation}

This expression is fitted to our numerical data by choosing the set of
parameters $\{C_1, C_2, e, \omega\}$ that minimize the
error 

\begin{equation}
E = \sum \left( \frac{P_B^{measured}-P_B^{fit}}{P_B^{measured}}\right)^2.
\end{equation}

A plot of $-log(P_B)/log(L)$ vs. $1/log(L)$ should then have an asymptotic ($L
\to \infty$) intercept equal to the leading exponent $e$. Similar fitting
procedures were used to produce Figures~\ref{fig:betaestimate} and 
\ref{fig:nuestimate}, where the leading exponent is $\beta/\nu$ and $1/\nu$
respectively. 

A fit of the data in Fig.~\ref{fig:components-all}b,d produces a rather
universal estimate $\beta'_{gr}/\nu = 0.22 \pm 0.02$.  In consequence the
rigid backbone is \emph{fractal} at $p_c$, with a fractal dimension $D_B=1.78
\pm 0.02$.  In the connectivity case (Fig.~\ref{fig:components-all}a), we
find~\cite{CFMbackbone} $\beta'/\nu =0.350 \pm 0.005$, or $D_B = 1.650 \pm
0.005$, which is consistent with the most precise prior work~\cite{RN,Gb}.

Now we consider $P_{\infty}$ and $P_D$.  In the connectivity case
(Fig.~\ref{fig:components-all}a), an analysis of the dangling ends and
infinite cluster probabilities (Fig.~\ref{fig:betaestimate}a) both lead to the
estimate $\beta/\nu =0.10-0.11$, in agreement with the exact result 5/48.  In
the rigidity case however, there are strong finite size effects and even at
sizes of $L=3200$ (joint-bar rigidity, Fig~\ref{fig:components-all}b), and
$L=4096$ (body-joint rigidity, Fig.~\ref{fig:components-all}d, it looks as
though the dangling probability may be saturating, while the infinite cluster
density continues to decrease.  Since $P_{\infty} = P_B + P_D$, it is expected
that asymptotically $P_{\infty}$ and $P_D$ must behave in the same manner.

Clearly the numerical results for the range of system sizes currently
available are still controlled by finite size effects, and the results depend
on the analysis method chosen.  Jacobs and Thorpe\cite{JaTh} chose to
interpret the infinite cluster probability as being key. A fit to the
$P_{\infty}$ data of Fig.~\ref{fig:components-all}b,d yields $\beta/\nu =
0.147 \pm 0.005$ (See Fig.~\ref{fig:betaestimate}b,c) in agreement with Jacobs
and Thorpe.  But a similar fit of the dangling end density gives $\beta/\nu
\sim 0.03$ for the joint-bar rigidity case (Fig.~\ref{fig:betaestimate}b) and
$\beta \sim 0.01$ for the body-joint rigidity case
(Fig.~\ref{fig:betaestimate}c).  In our previous work\cite{MDlet2} we were
guided by the Cayley tree results~\cite{Cayley} which indicated a first order
jump in the infinite cluster probability.  We thus chose to interpret
Fig.~\ref{fig:components-all}b,d as indicating a saturation of the infinite
cluster probability at the dangling end value of about $0.1$.  Having extended
our data from $L=1024$ to $L=4096$, it now looks more likely that a small
value of $\beta$ occurs in the rigidity case (Fig.~\ref{fig:betaestimate}b,c),
though much larger simulation sizes are required to find $\beta/\nu$
precisely.

Due to the slow finite size effects found in the analysis of the infinite
cluster and dangling end probabilities, it is natural to be concerned about
the effect of boundary conditions and other, usually non-universal, parameters
on the observed results.  For generic joint-bar rigidity case on triangular
lattices we thus tested a variety of different boundary conditions for both
site and bond dilution.  This data is presented in
Fig.~\ref{fig:all-rig-cases}, from which it is seen that the conclusions drawn
from the case of rigidity percolation with applied bus bars are quite robust.

\vbox{
\begin{figure}[htb]
\vbox{
\centerline{\psfig{figure=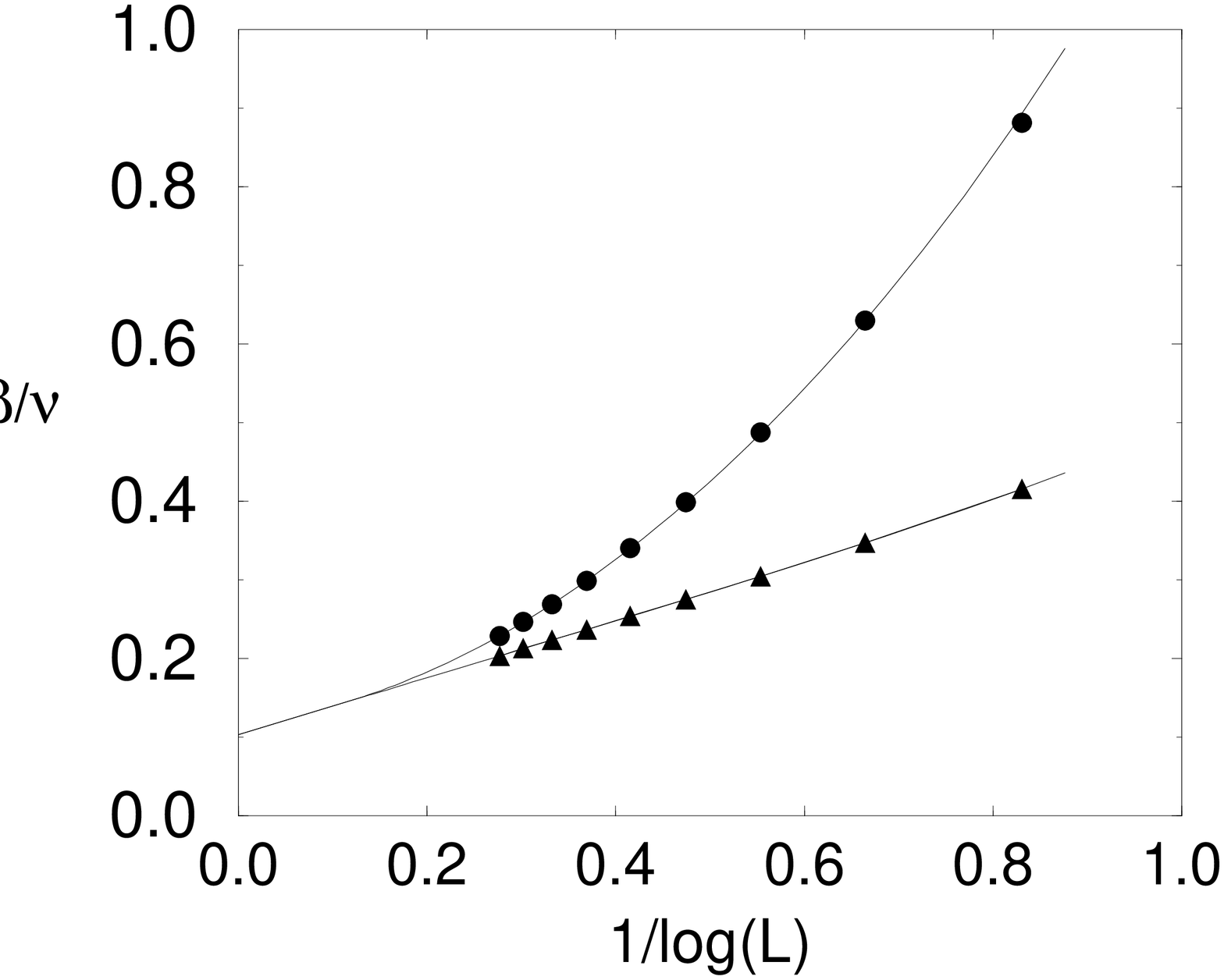,width=6.8cm,angle=270}}
\centerline{\psfig{figure=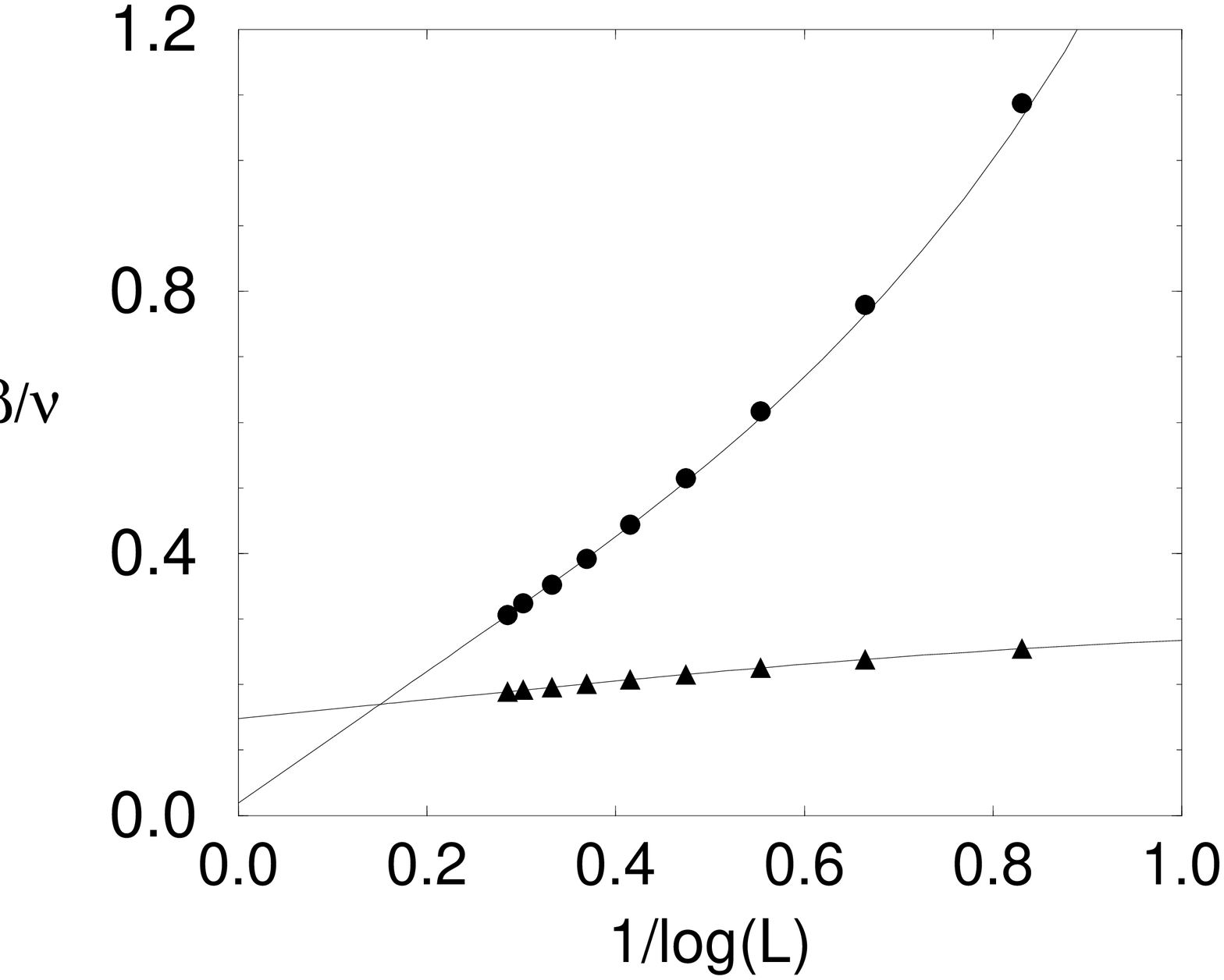,width=6.8cm,angle=270}} 
\centerline{\psfig{figure=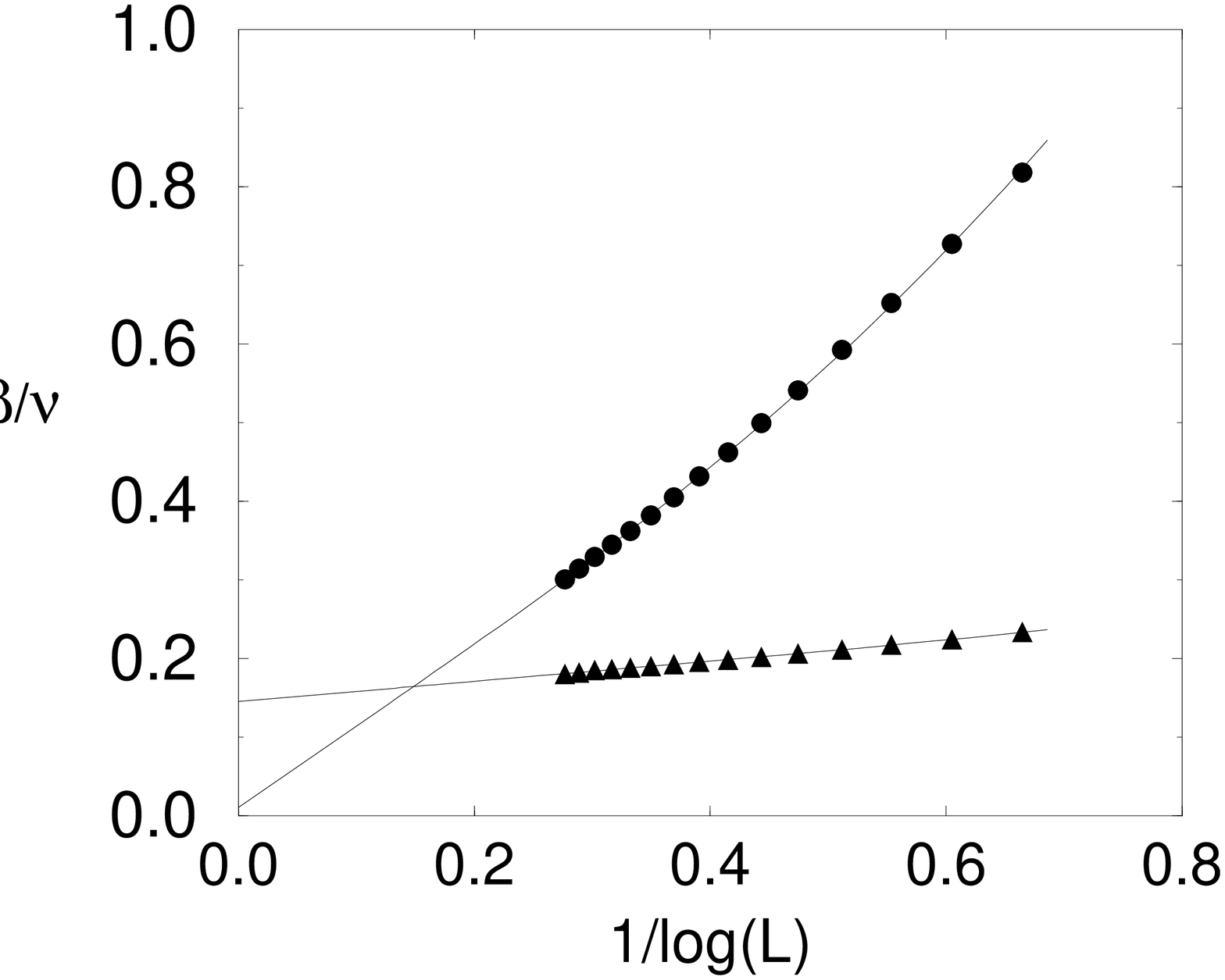,width=6.8cm,angle=270}} 
}
\vskip 1cm
\caption{ 
  The spanning cluster density exponent $\beta/\nu$ as numerically estimated
  for {\bf a)} connectivity percolation on a square lattice ($L^{max}=4096$),
  {\bf b)} rigidity percolation on a triangular lattice ($L^{max}=3200$) and
  {\bf c)} body rigidity on a square lattice ($L^{max}=4096$).  Two estimates
  result in each case from fitting the scaling of spanning cluster density
  (triangles) and dangling end density (circles).  Solid lines are fits using
  \hbox{Eq. \protect (\ref{eq:fit})} }
\label{fig:betaestimate}
\end{figure}
}

Finally, the behavior of the dangling end density as a function of $p$ is also
quite striking.  This data is presented in Fig.~\ref{fig:blue_vsp}.  At very
high $p$, nearly all bonds belong to the backbone, so the dangling end density
approaches zero. Below $p_c$, there is no infinite cluster,
 so there are again no dangling
ends.  There then must be a maximum in the density of dangling ends between
$p_c$ and $p=1$.  As seen in Fig.~\ref{fig:blue_vsp}, the interesting feature
is the abrupt drop in the dangling density at $p_c$, a feature that appears to
become more pronounced with increasing sample size.  It is tempting to
interpret this as definitive evidence of a first order rigidity transition,
but it is also consistent with the strong finite size effects seen in
Figs.~\ref{fig:components-all}b,d and Fig. 6, so we must await large lattice
simulations for a definitive analysis.

\vbox{
\begin{figure}[htb]
\centerline{\psfig{figure=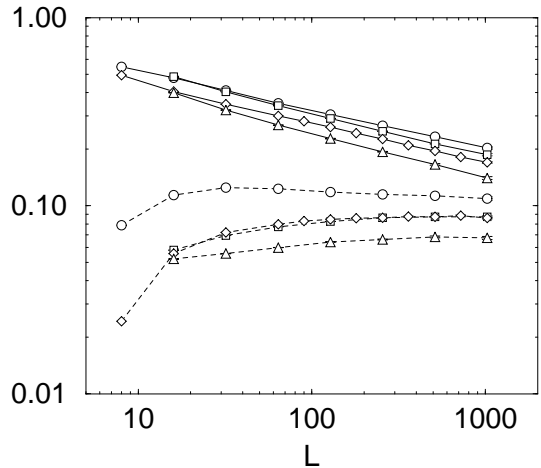,width=8cm,angle=270}} 
\vskip 0.5cm
\caption{ 
Backbone density $P_B$ (solid lines) and dangling-end density $P_D$ (dashed
lines)  as a function of sample
size at the percolation threshold of each sample, on triangular lattices for:
bond-diluted AS with bus-bars (circles), site-diluted AS with bus-bars
(diamonds), site-diluted AS without bus-bars (triangles) and site-diluted IS
without bus-bars. The AS percolation point without bus-bars is defined as the
concentration of sites or bonds for which there is for the first time a rigid
connection between at least one pair of points on opposite sides of the
sample. 
}
\label{fig:all-rig-cases}
\end{figure}
}

\begin{figure}[htb] \vbox{ 
\centerline{ \psfig{figure=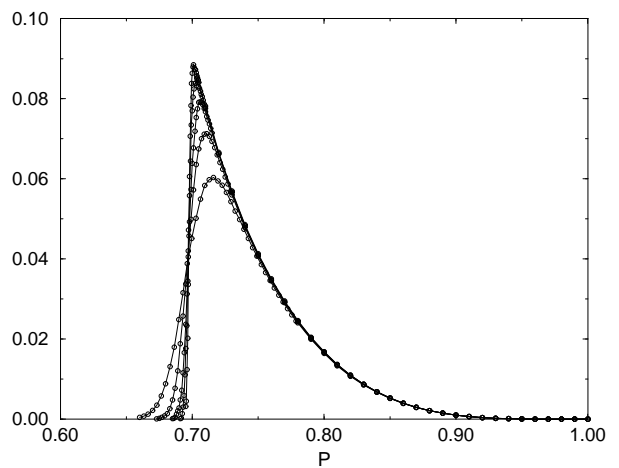,width=8cm,angle=270} }
\vskip 0.5cm
\caption{Fraction of dangling ends on the ($g=2,G=3$) generic rigidity
infinite cluster, as a function of
$p$, for site-diluted triangular lattices of size $L=$ 32, 64, 128, 256, 512
and 1024. Data shown here are for the AS case with bus-bars.  }
\label{fig:blue_vsp}
} \end{figure}

Now we turn to the calculation of the correlation length exponent for rigidity
percolation.  When there is a second-order rigidity transition, there is a
diverging correlation length $\xi \sim |p-p_c|^{-\nu}$.  We can find the
exponent $\nu$ of this divergence by measuring the sample-to-sample
fluctuations in $p_c$ as a function of $L$. The dispersion \hbox{$\sigma(L) =
  \sqrt( < p_c^2 >_L - < p_c >_L^2 )$}, and according to finite-size-scaling
$\sigma(L) \sim L^{-1/\nu}$.
\vbox{
\begin{figure}[htb]
\vbox{
\centerline{\psfig{figure=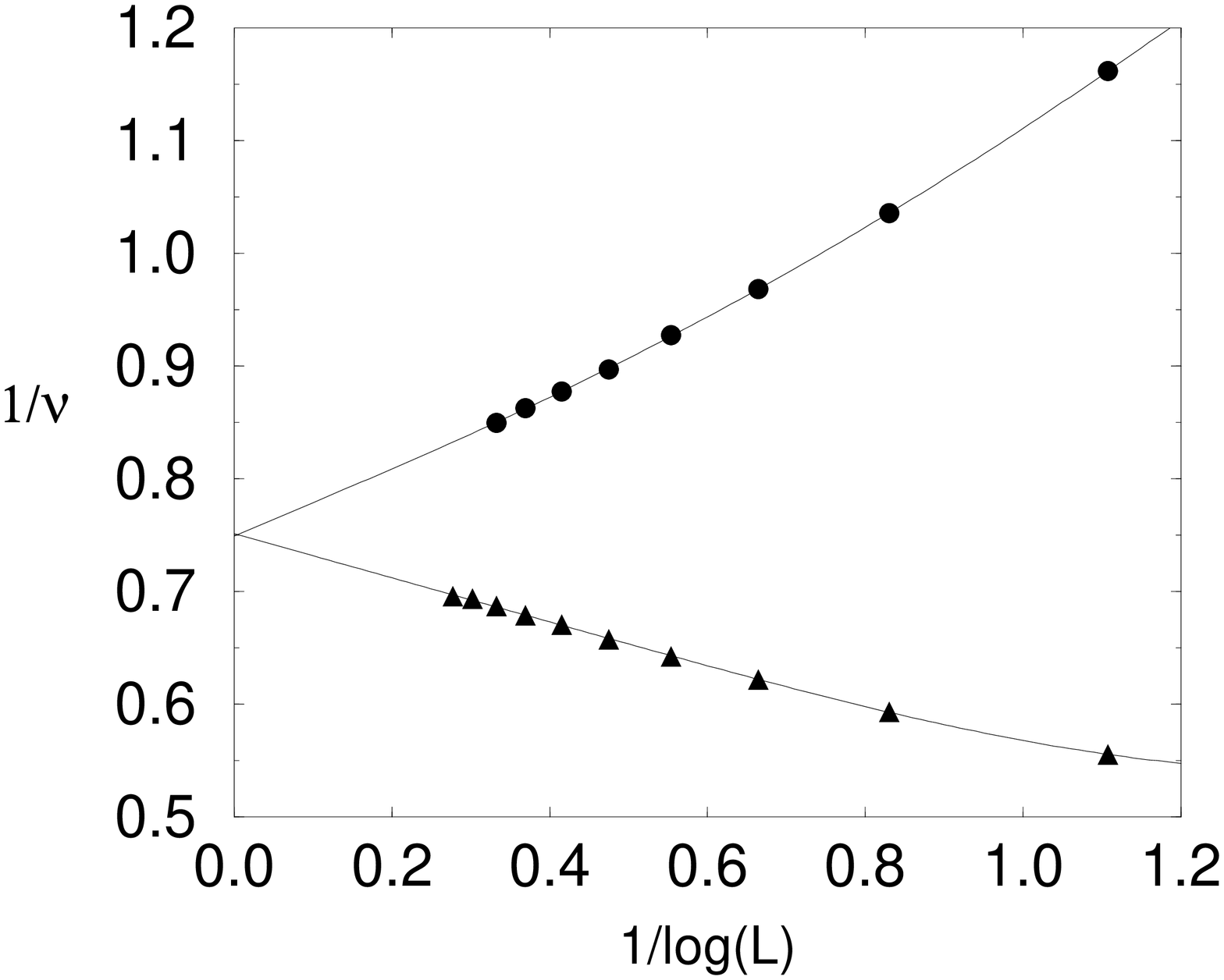,width=7cm,angle=270}}
\centerline{\psfig{figure=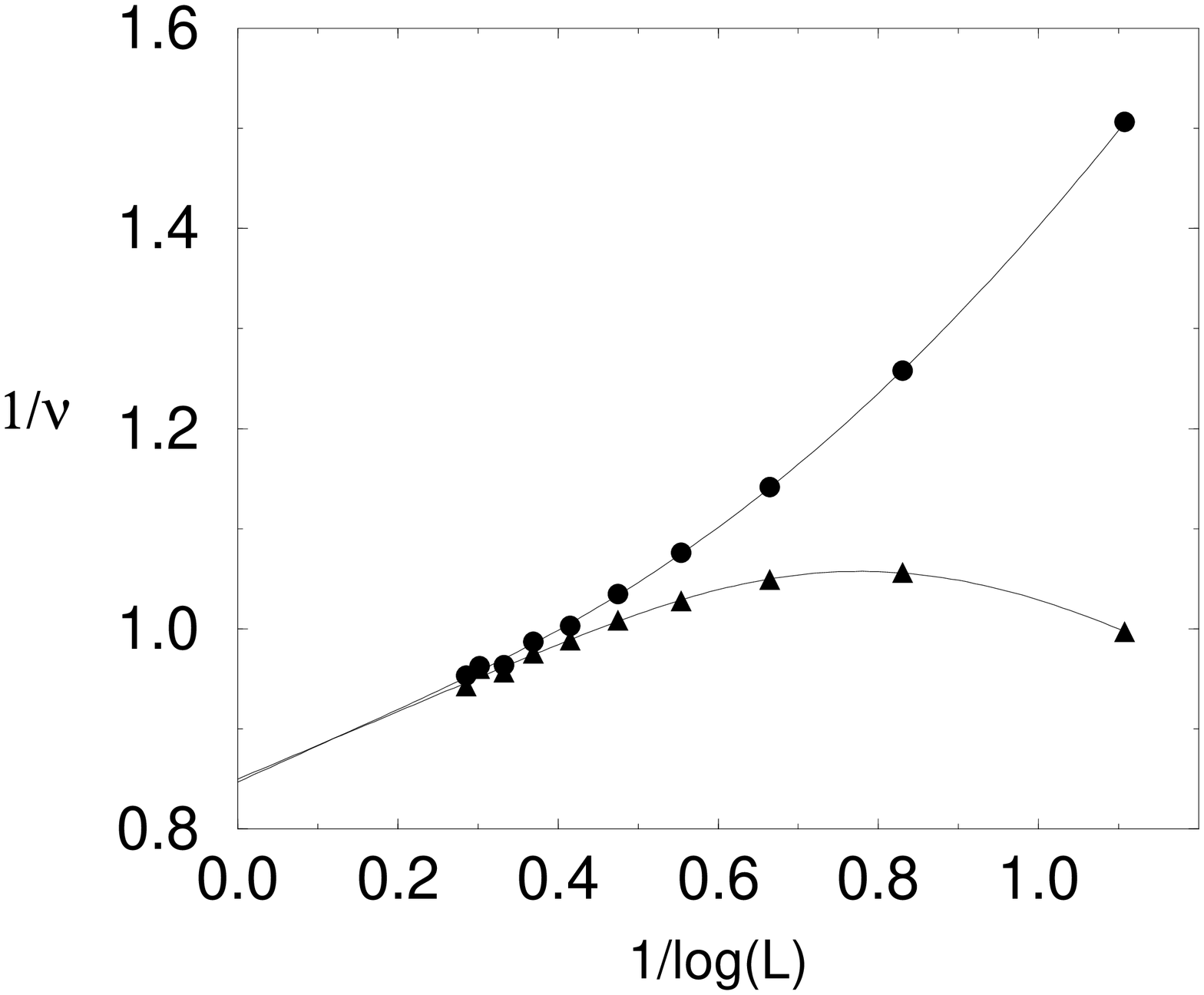,width=7cm,angle=270}} 
\centerline{\psfig{figure=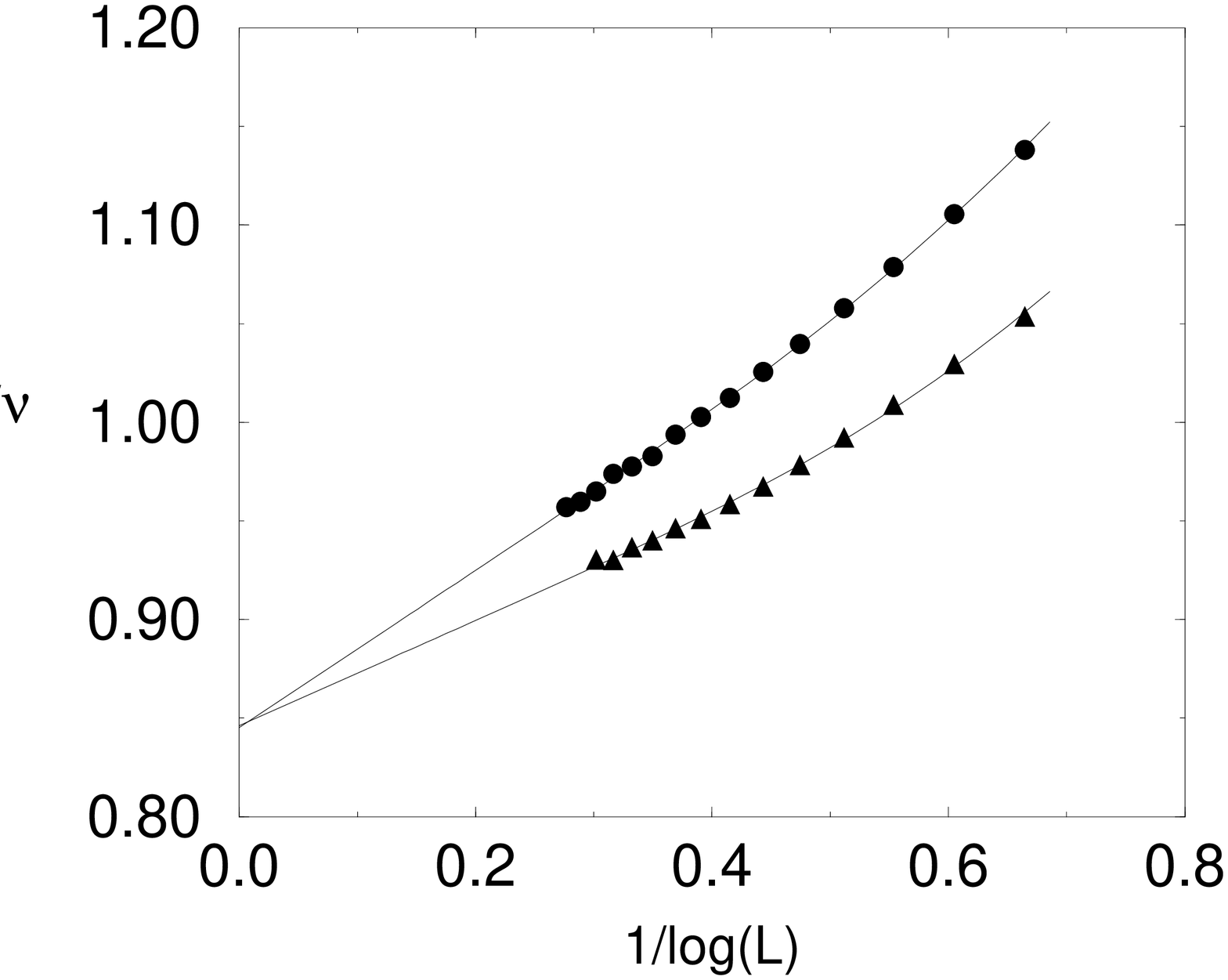,width=7cm,angle=270}} 
}
\vskip 0.5cm
\caption{ 
The thermal exponent $1/\nu$ as numerically estimated for {\bf
  a)} connectivity percolation ($g=G=1$) on square lattice, {\bf b)} rigidity
percolation ($g=2, G=3$) on a triangular lattice and {\bf c)} body rigidity on
a square lattice ($G=g=3$). Two independent estimates result in each case
from fitting the scaling of red bonds (triangles) and fluctuations 
in $p_c$ (circles). 
}
\label{fig:nuestimate}
\end{figure}
}

\begin{figure}[htb] 
\vbox{
\centerline{\psfig{figure=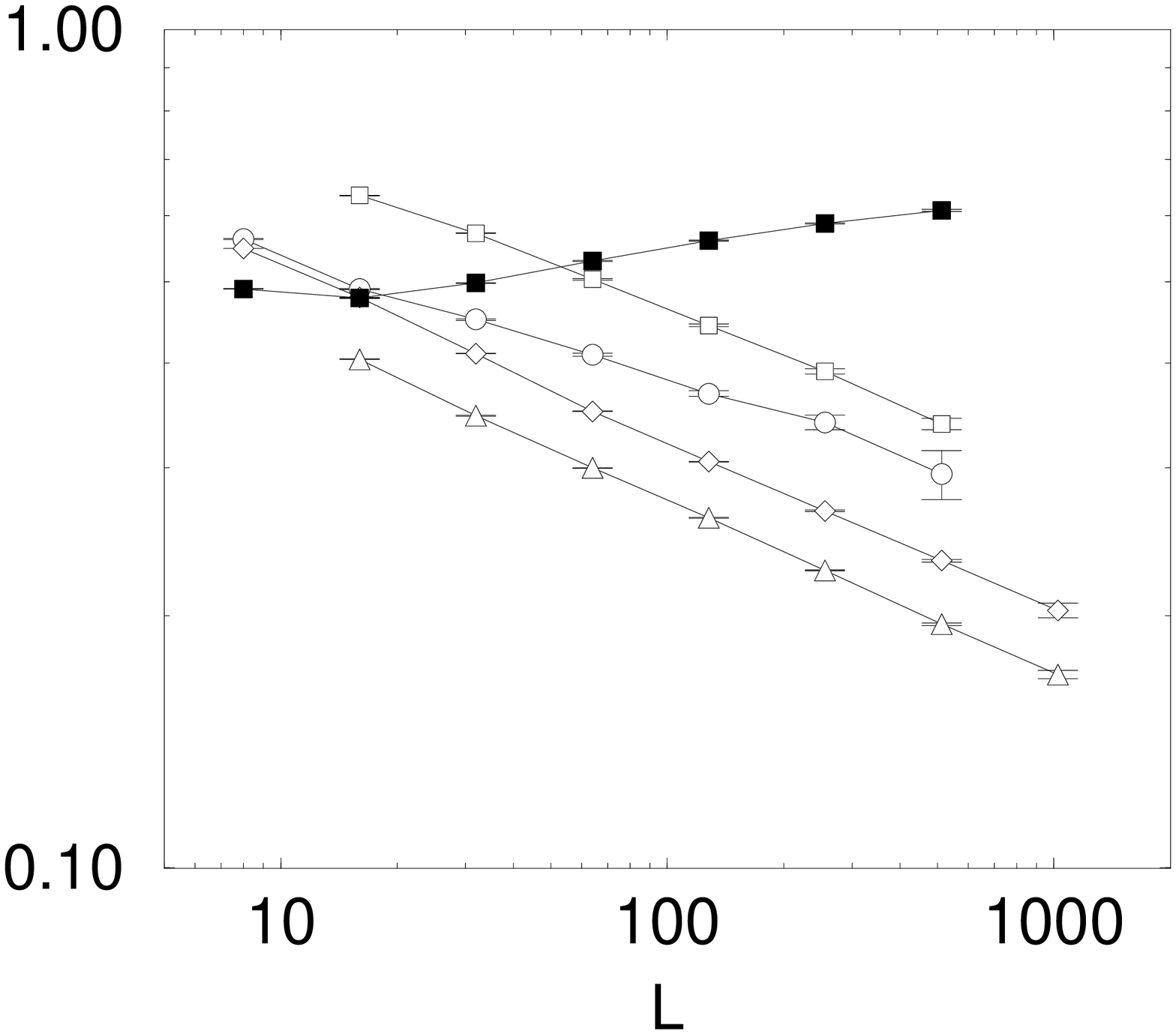,width=7cm,angle=270}} 
\centerline{\psfig{figure=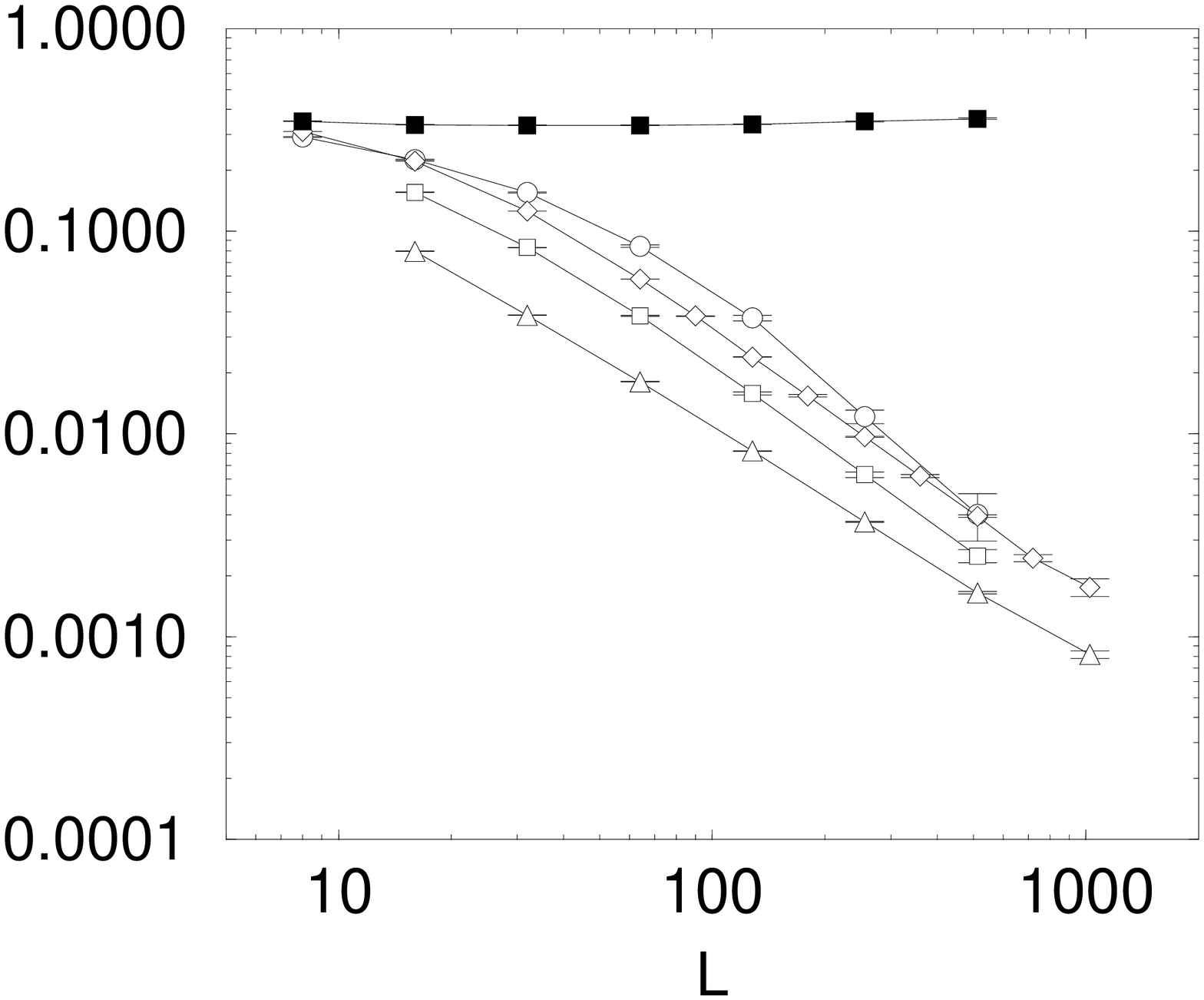,width=7cm,angle=270}} 
\centerline{\psfig{figure=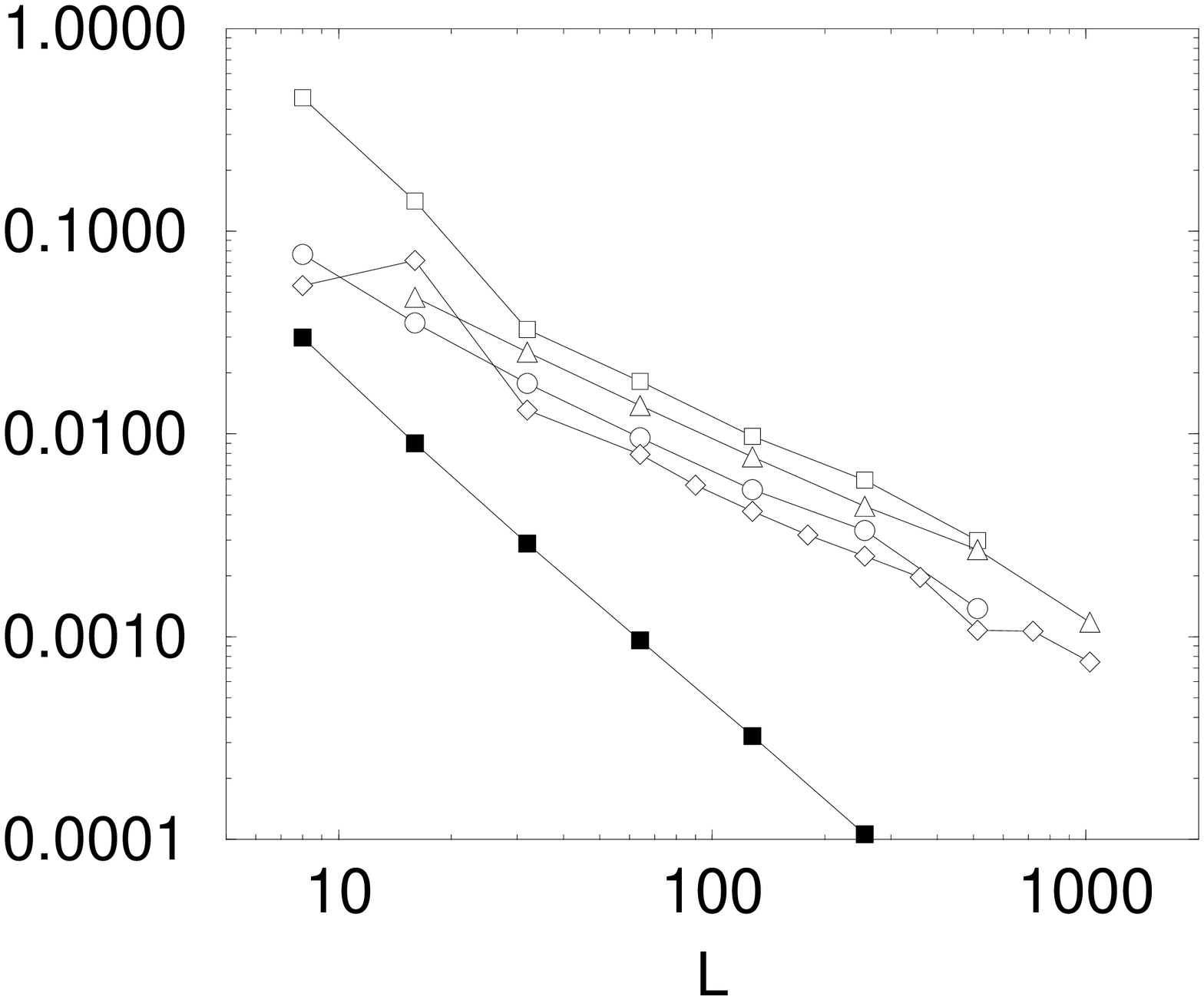,width=7cm,angle=270}}  
\vskip 0.5cm
\caption{ 
The volume fraction of {\bf a)} backbone bonds {\bf b)} cutting bonds, 
and {\bf c)} the fluctuation $\sigma(p_c)$ at the rigidity
threshold for: Square lattice with $q=0$ - this is the
braced square net (filled squares); Square lattice with
$q=0.1$ (circles); Square lattice with $q=0.40$ (squares); Triangular lattice
with bond dilution (diamonds) and Triangular 
lattice with site dilution (triangles).
} 
\label{fig:compare}  
} \end{figure}

An asymptotic analysis for $\sigma(L)$ is shown in Fig.~\ref{fig:nuestimate}a
for connectivity percolation, in \ref{fig:nuestimate}b for joint-bar rigidity
and in \ref{fig:nuestimate}c for body-joint rigidity.  From these figures we
estimate $1/\nu=0.75 \pm 0.01$ (the exact value is $1/\nu = 3/4$) for
connectivity percolation and $1/\nu = 0.85 \pm 0.02 $ for rigidity
percolation.  This provides further strong evidence that rigidity percolation
is second order in two dimensions, though {\it not} in the same universality
class as scalar percolation.  In the case of the first order rigidity on the
braced square net (Fig.~\ref{fig:compare}c), the variations in $p_c$ behave as
$L^{-3/2}$, in accordance with analytical results for this
model~\cite{CFMtbp}.

Our algorithm also identifies the cutting (also called {\it red} or critical)
bonds at the percolation point, for the case of AS percolation.  The number
$N_R$ of red bonds scales at $p_c$ as $L^{x}$.  Coniglio~\cite{Coniglio} has
shown that $x=1/\nu$ exactly, for \emph{scalar percolation}. Numerical
evidence suggesting that $x=1/\nu$ also in rigidity percolation was first
presented in~\cite{MDlet1}. It is in fact possible to extend Coniglio's
reasoning to the case of central-force rigidity percolation~\cite{CFMtbp}. It
turns out that $x = 1/\nu$ has to be rigorously satisfied also in this case,
and therefore $\sigma(L)$ and $1/N_R(L)$ must have the same slope in a log-log
plot.  Analysis of the number of cutting bonds is also presented in
Fig.~\ref{fig:nuestimate}, and yields values of $1/\nu$ consistent with the
analysis of variations in percolation thresholds described in the previous
paragraph.
  
Since the Cayley tree model~\cite{Cayley} gives behavior quite similar to the
braced square net~\cite{Obkv}, i.e. a first-order rigidity transition, it is
interesting to ask whether the rigidity transition is ``usually'' like that on
the braced square net (i.e.  first order), or whether the second order
transition found on triangular lattices is more typical.  In order to probe
this issue, we analyze a model which interpolates between the braced square
net and the triangular lattice.  In the braced square net, the random
diagonals are present with probability $p_d$, to make the lattice rigid it is
sufficient (though not necessary) to add one diagonal to every row of the
square lattice.  The probability that a spanning cluster exists is then
\hbox{$P_+ = (1-(1-p_d)^L)^L$}, from which we find $p_{d*} \sim lnL/L$.

We generalize this model by randomly adding the diagonals (with probability
$p_d$) to a square lattice whose bonds have been diluted with probability $q$.
The braced square net is $q=0$, while if $q=1-p_d$ this model is equivalent to
the bond-diluted triangular lattice.  Typical results for various values of
$q$ are presented in Fig.~\ref{fig:compare}. It is seen that even for a small
amount of dilution of the square lattice, e.g. $q=0.10$, the rigidity
transition returns to the behavior characteristic of the homogeneously diluted
triangular case (see Fig.~\ref{fig:compare}).  We find that for sufficiently
large lattice sizes, the universal behavior found in the other rigidity cases
holds for any finite $q<0.5$ (for larger values of $q$ it is not possible to
rigidize the lattice by randomly adding diagonals), and we suggest that the
``fully-first-order'' transition (i.e. a first-order backbone) only occurs in
the special case of a perfect (undiluted) square lattice.

\section{Conclusions} 
\label{sec:conclusions} 

We have compared three types of percolation transition in two dimensions: the
connectivity transition and; the generic rigidity transition on the triangular
lattice and; the generic rigidity transition on the braced square lattice.  A
summary of our understanding is as follows: (i) The generic rigidity
transition on triangular lattices is second order with $\nu=1.16\pm 0.03$,
$0\le\beta \le 0.2$, $\beta' = 0.25\pm 0.02$ and; (ii) The rigidity transition
on the braced square net is first order with finite backbone, spanning cluster
and cutting bond densities at the percolation threshold.  Only the value of
$\beta$ for the generic rigidity transition on triangular lattices remains
controversial, due to the very strong finite size effects in that case.

To illustrate the fact that our data is inconsistent with a first order
backbone in the site-joint rigidity case, we have developed the following
scaling argument.
  
Assume that the backbone mass\cite{count} scales as $M_B \sim L^{D_B}$ at
$p_c$. If the backbone is compact then $D_B=d$, the dimension of the system.
The backbone mass is composed of red (or cutting) bonds plus ``blobs'' of
overconstrained, or self-stressed bonds (See Fig.~\ref{fig:spcluster}).
Therefore $M_B = M_{red} + M_{blobs}$. The number of red bonds in the backbone
scales as $M_{red} \sim L^{1/\nu}$, as analytical
results~\cite{Coniglio,CFMtbp} and the simulations reported here show. Let us
furthermore write $M_{blobs}= n_{blobs} \times m_{blobs}$ where $n_{blobs}$ is
the number of blobs in the backbone, and $m_{blobs}$ be the average number of
bonds in a blob. Therefore \hbox{$L^{D_B} \sim L^{1/\nu} + n_{blobs}
  m_{blobs}$}.  Now, the AS backbone is an \emph{exactly isostatic} body-bar
structure, formed by rigid clusters (blobs) joined by bars (red bonds) so that
counting of degrees of freedom is exact on it and so $M_{red} = 3 n_{blobs} +
2 n_s -3$~\cite{CFMalg}. Here $n_s$ is the number of sites in the backbone,
that do not belong to a blob (see Fig.~\ref{fig:spcluster}).  This identity is
known as Laman's condition\cite{Laman,CFMalg}, and results from the fact that
each red bond acts as a bar and therefore restricts one degree of freedom,
while each blob has three degrees of freedom and isolated sites have two. The
backbone is a rigid cluster and therefore has three overall degrees of
freedom. We do not need to know $n_s$ for our argument. It is enough to notice
that $n_{blobs} \leq M_{red}/3 \sim L^{1/\nu}$. We can thus write

\begin{equation}
L^{D_B} \leq L^{1/\nu} ( 1 + m_{blobs}/3 )
\end{equation}   

To this point, we have made no assumption about the character (compact or
fractal) of the backbone, so it is valid in general.  If the transition is
second-order, there is a divergent length (for example, the size of rigid
clusters), and we expect $m_{blobs}$ to diverge with system size. Therefore a
non-trivial value results for $D_B$, as we find numerically. If on the other
hand there is no diverging length in the system, then $m_{blobs} \to $
constant for large systems and $D_B = d = 1/\nu$ exactly.  We thus see that a
compact backbone requires an extensive number of cutting bonds, and this in
turn can only be satisfied if $\nu = 1/d$ exactly\cite{aclaracion}. This is
completely inconsistent with our data, and so the possibility of a first order
backbone is remote in two dimensions.

The first order rigidity transition exhibited by the braced square net seems
to be atypical as we illustrated using a model which tunes continuously from
that limit toward the generic triangular lattice.  We found that even a small
deviation from the braced square lattice limit leads to a behavior similar to
that of the triangular lattice.  It would be intriguing if there were a
tricritical point at which first order rigidity ceases and second order
rigidity sets in, but we have not found a model which exhibits that behavior.
Nevertheless there are a large number of other rigidity models in two
dimensions, so the possibility is not yet ruled out.
   
C.~M. acknowledges financial support of CNPq and FAPERJ, Brazil. This work has
been partially supported by the DOE under contract DE-FG02-90ER45418, and  the
PRF. We are grateful to HLRZ J\"ulich for the continued use of their
computational resources. 



\begin{references}
\bibitem{infrig}
  
  We refer in this paper to ``infinitesimal rigidity'' as rigidity for short.
  Infinitesimal rigidity is a stronger condition than rigidity, and implies
  that no transformation, even an infinitesimal one, leaves all bar lengths
  invariant.  See e.g. Refs.~\cite{Rig_def,Hendr,CFMalg}
\bibitem{count}
  
  We define the mass of the backbone as the number of bonds belonging to it.
  It is not correct to count sites in rigidity percolation, since they can
  simultaneously belong to several rigid clusters.
\bibitem{Maxap}
  
  The Maxwell approximation consists in ignoring the existence of redundant
  bonds, i.e.  assuming that each present bond eliminates one degree of
  freedom from a total of $F=gN$. The critical concentration is then obtained
  by equating $f=F/N$ to zero.
\end{references}
\end{document}